\documentclass[sigplan,10pt]{acmart}
\setcopyright{none}
\renewcommand\footnotetextcopyrightpermission[1]{} 
\setcopyright{none}
\renewcommand\footnotetextcopyrightpermission[1]{} 
\AtBeginDocument{%
  }

\usepackage[normalem]{ulem}
\usepackage{hyperref}
\usepackage{listings}
\usepackage{graphicx}
\usepackage{tablefootnote}
\usepackage{caption}
\usepackage{subcaption}
\usepackage{multirow}

\begin{document}

\title{Viewing Allocators as Bin Packing Solvers Demystifies Fragmentation}

\author{Christos P. Lamprakos}
\email{cplamprakos@microlab.ntua.gr}
\orcid{0000-0002-3370-857X}
\affiliation{%
  \institution{National Technical University of Athens}
  \streetaddress{9, Iroon Polytechniou St.}
  \city{Athens}
  \country{Greece}
  \postcode{157 80}
}
\affiliation{%
  \institution{Katholieke Universiteit Leuven}
  \streetaddress{Oude Markt 13}
  \city{Leuven}
  \country{Belgium}
  \postcode{3000}
}

\author{Sotirios Xydis}
\orcid{0000-0003-3151-2730}
\email{sxydis@microlab.ntua.gr}
\affiliation{%
  \institution{National Technical University of Athens}
  \streetaddress{9, Iroon Polytechniou St.}
  \city{Athens}
  \country{Greece}
  \postcode{157 80}
}

\author{Francky Catthoor}
\orcid{0000-0002-3599-8515}
\email{francky.catthoor@imec.be}
\affiliation{%
  \institution{IMEC Science Park}
  \streetaddress{Gaston Geenslaan 14}
  \city{Leuven}
  \country{Belgium}
  \postcode{3001}
}
\affiliation{%
  \institution{Katholieke Universiteit Leuven}
  \streetaddress{Oude Markt 13}
  \city{Leuven}
  \country{Belgium}
  \postcode{3000}
}

\author{Dimitrios Soudris}
\orcid{0000-0002-6930-6847}
\email{dsoudris@microlab.ntua.gr}

\affiliation{%
  \institution{National Technical University of Athens}
  \streetaddress{9, Iroon Polytechniou St.}
  \city{Athens}
  \country{Greece}
  \postcode{157 80}
}

\renewcommand{\shortauthors}{Lamprakos et al.}

\begin{abstract}
The main goal of dynamic memory allocators is to minimize memory fragmentation. Fragmentation results from the interaction of workload behavior and allocator policy. There are, however, no works systematically capturing this interaction in an informative data structure. We consider this gap responsible for the absence of a standardized, quantitative fragmentation metric, the lack of workload characterization techniques with respect to their dynamic memory behavior, and the absence of an open, widely used benchmark suite targeting dynamic memory allocation. Such shortcomings are profoundly asymmetric to the operation's ubiquity.

This paper presents a trace-based simulation methodology for constructing representations of workload-allocator interaction. We use two-dimensional rectangular bin packing (2DBP) as our foundation. Classical 2DBP algorithms minimize their products' makespan, but virtual memory systems employing demand paging deem such a criterion inappropriate. We view an allocator's placement decisions as a solution to a 2DBP instance, optimizing some unknown criterion particular to that allocator's policy. Our end product is a compact data structure that fits e.g. the simulation of 80 million requests in a 350 MiB file. By design, it is concerned with events residing entirely in virtual memory; no information on memory accesses, indexing costs or any other factor is kept.

We bootstrap our contribution's significance by exploring its relationship to maximum resident set size (RSS). Our baseline is the assumption that less fragmentation amounts to smaller peak RSS. We thus define a fragmentation metric in the 2DBP substrate and compute it for 28 workloads linked to 4 modern allocators. We also measure peak RSS for the 112 resulting pairs. Our metric exhibits a strong monotonic relationship (Spearman coefficient $\rho>0.65$) in half of those cases: allocators achieving better 2DBP placements yield $9\%$-$30\%$ smaller peak RSS, with the trends remaining consistent across two different machines. 

Considering our representation's minimalism, the presented empirical evidence is a robust indicator of its potency. If workload-allocator interplay in the virtual address space suffices to evaluate a novel fragmentation definition, numerous other useful applications of our tool can be studied. Both augmenting 2DBP and exploring alternative computations on it provide ample fertile ground for future research.
\end{abstract}



\maketitle
\pagestyle{plain}

\section{Introduction}
In their seminal 1995 survey, Wilson et al. contributed both a comprehensive taxonomy and a grounded critique of dynamic storage\footnote{We use "storage" instead of "memory" as a tribute to the outstanding survey by Paul R. Wilson et al.~\cite{wilson1995surv} The matter at hand will at all times remain non-moving virtual memory allocation.} allocation (DSA)~\cite{wilson1995surv}, noting the inherent difficulty in defining fragmentation, the inadequacy of basing designs on synthetic workloads, and the lack of novelty in new allocator policies. To this day, we have not converged to a single, measurable definition of fragmentation~\cite{realfrag}, neither do we possess a principled method for workload characterization--despite the fact that program behavior partly controls fragmentation.

Most noticeably, there is no standardized memory allocation benchmark suite. Motivation sections often adopt synthetic test cases~\cite{snmalloc} even though we know such practices to be inadequate. Applications used for evaluation are selected on intuitive grounds of being "dynamic enough". Certain classes, such as database and web browsing workloads, are preferred over others with no proper justification. Worse, "internal" workloads are at times used~\cite{learnalloc}, obstructing transparency and reproducibility. 

To define fragmentation, functions of resident set size (RSS) are employed~\cite{mesh}, this being probably an influence from the four alternative formulations proposed in~\cite{fragsolved}. No attempts have been made to evaluate the utility of each option, in spite of the original authors emphasizing the value of such an investigation. Whether general-purpose policies suffice to handle modern workloads or not is unclear, since works supporting both views exist~\cite{reconsider, learnalloc}. Meanwhile, novel allocator designs keep appearing, server workloads are switching to larger page sizes, persistent memory has entered the picture~\cite{persistence} and the world relies on massively multitasking, resource-sharing datacenters more each passing year.

Our motivation is formed by the hidden cost imposed to systems from the aforementioned gaps, and amplified by the ubiquitous and indispensable nature of DSA. This paper is built on the conjecture that \textit{returning to first principles is necessary} if a rigorous memory management theory is to be established. By first principles, fragmentation is the main enemy of any allocator, and it is a function of the interaction between workload behavior and allocator policy~\cite{wilson1995surv}. Our research objective is thus to find a \textit{structured representation} capturing said interaction, and a systematic approach which will enable this in practice for realistic workloads.

We start by observing two distinct branches of DSA research: practical work intended to operate on realistic environments, and theoretical work exploring limits and other aspects of allocator policy. For instance, best fit and first fit policies were assigned their worst case costs by Robson in 1977~\cite{robson1977worst}. More recently, Appel and Naumann provided a verification procedure for sequential \texttt{malloc} and \texttt{free} operations~\cite{appel2020verif}. 

Of particular interest to us is the noted resemblance of DSA to a variation of two-dimensional rectangular bin packing (2DBP)~\cite{chrobak1988some, buchsbaum}. Up to this point 2DBP has been treated as an NP-hard optimization problem~\cite{garey_jognson_1979}, with existing approximate algorithms generating placements of minimal makespan. We do not intend to create a novel 2DBP algorithm; optimizing virtual memory makespan is meaningless in the context of demand paging. But what if we viewed \textit{allocators themselves} as 2DBP "algorithms" with unknown optimization criteria? The resulting structures, i.e. sets of rectangles defined by a program's dynamic memory requests sequence and placed according to an allocator's policy, should multiplex enough of the workload-allocator interaction that we are targeting. 

To this end, we present a systematic, portable, trace-based simulation methodology for representing workload-allocator interaction as 2DBP instances. To conclude whether our produced representations contain any information of practical value, we investigate their relationship to maximum resident set size (RSS). More specifically, we define fragmentation as the ratio between gaps and used memory in the 2DBP space, and measure it for 28 workloads linked to 4 modern allocators. For $46.4\%$ of the studied workloads, 2DBP-based fragmentation and maximum RSS exhibit a strong monotonic relationship as dictated by Spearman's correlation coefficient ($\rho>0.65$). Lower fragmentation in 2DBP yields up to $30\%$ smaller memory footprint in the real world, with the trends remaining consistent across two different machines. Our contributions can thus be summarized as:

\begin{itemize}
    \item a novel perspective emphasizing the need for a principled study of workload-allocator interaction, motivated by important gaps identified in the state-of-the-art
    \item a widely applicable methodology for constructing 2DBP representations of arbitrary workloads and non-moving allocators
    \item a first empirical study of 2DBP's informational content
    \item a novel definition of external memory fragmentation
    \item an extensive discussion on our results' implications for DSA, motivating the next steps to be taken on top of our foundation
\end{itemize}

In Section \ref{sec:bck} we elaborate on our representation's derivation, as well as on the notion of 2DBP-based fragmentation. In Section \ref{sec:meth} we accurately describe the mechanisms implemented to actualize our methodology. We present our results in Section \ref{sec:res} and discuss their implications in Section \ref{sec:disc}. Related work is presented in Section \ref{sec:rw}, and Section \ref{sec:end} closes the main text with an overview of our conclusions. Potential enhancements of the described infrastructure as well as openings for future research are, last but not least, the matter at hand of Section \ref{sec:lims}.

\section{Background}
\label{sec:bck}
Dynamic memory allocators receive a series of requests from the programs they are linked to. Two main request types exist: \textit{allocation} of $n$ bytes and \textit{deallocation} of a previously allocated block. 

Real allocation requests come in several variations. A program may need memory blocks of specific alignment, or blocks initialized as a zero-valued array. The program may even ask for an allocated block to be resized. Upon successful allocation, a pointer to the newly acquired memory is returned to the program. Deallocation requests are, on the other hand, straightforward. The program informs the allocator via a previously obtained memory pointer that it does not need the corresponding block any more.

An allocator's decisions on block placement and free memory management form its \textit{policy}. On the program side, the distribution of allocation sizes requested as well as the particular sequence of requests jointly form its \textit{behavior}. The goal of a good policy is to minimize \textit{fragmentation}, which means to waste minimal amounts of extra memory beyond what the program actually requested. Two types of fragmentation exist, namely internal and external. Internal fragmentation treats wasted memory within blocks (i.e. returning more bytes than requested); external fragmentation focuses between blocks (e.g. putting blocks that die together in non-consecutive places). In this paper we are concerned exclusively with external fragmentation, which is the hardest to deal with--even though our methodology can be easily extended to account for the internal kind as well. External fragmentation is a function of the \textit{interaction} between allocator policy and program behavior~\cite{wilson1995surv}. Several definitions have been proposed over the years~\cite{fragsolved, realfrag}.

A 2DBP instance comprises a series of \textit{unplaced} blocks in the form of $(start, end, height)$ tuples. An acceptable \textit{solution} to 2DBP is a placement with no overlapping blocks. For the purposes of our paper there is no need to distinguish between placed and unplaced blocks, so with the term "2DBP" we refer both to the requests \textit{and} the allocator's responses to each request (all blocks are already placed by the time we depict them). The concepts involved are best described by example. Let us consider the below requests sequence:

\begin{enumerate}
    \item \texttt{A = malloc(1)}
    \item \texttt{B = malloc(2)} 
    \item \texttt{free(A)}
    \item \texttt{C = malloc(3)}
    \item \texttt{free(B)}
    \item \texttt{free(C)}
\end{enumerate}

\begin{figure}[t!]%
    \centering
    \includegraphics[width=\columnwidth]{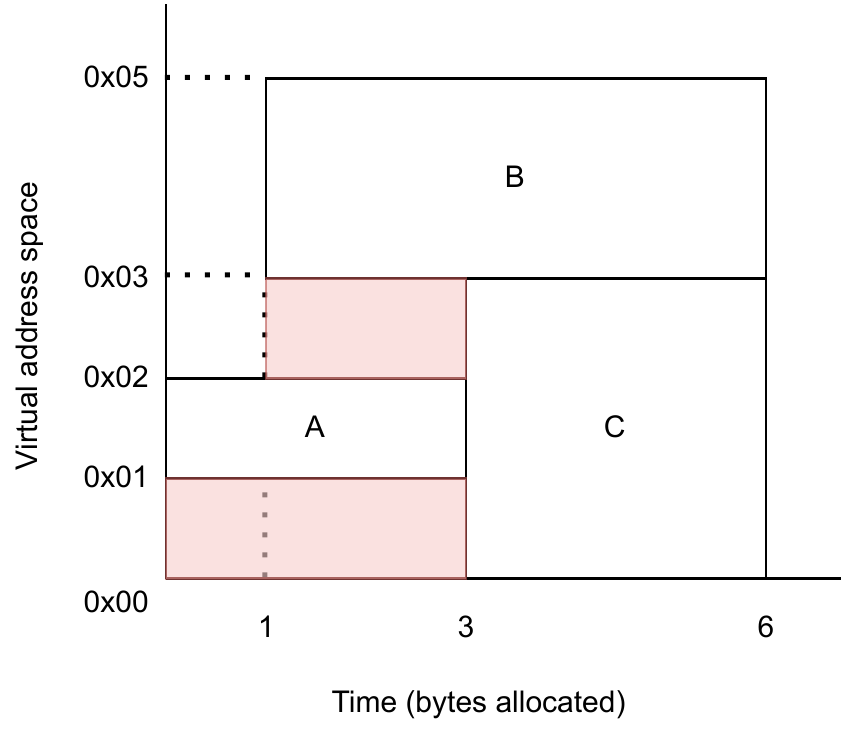}
    \caption{A simple 2DBP example.}%
    \label{fig:frag}
    \vspace*{-3.5mm}%
\end{figure}

Figure \ref{fig:frag} combines these requests with an imaginary allocator's responses, placing block A at virtual address 0x01, block B at 0x03 and block C at 0x00. The figure's horizontal axis measures time in allocated bytes. Time progresses forward after each allocation request, and remains unaltered after each deallocation request. It is important to differentiate between a typical memory block, for example ten contiguous bytes of virtual memory, and a 2DBP block like C above, which is a virtual memory block of three bytes occupied for a duration of three allocated bytes. We will be referring to the first kind as "blocks" and to the second kind as "jobs".

Normally 2DBP algorithms optimize a placement's make-span, meaning the total size of the address range used (in Figure \ref{fig:frag} the makespan equals to 5). We have already emphasized that in the scope of this paper, the allocators are the ones producing the placements; we are merely recording their decisions \textit{as if} they were solving a 2DBP problem. Consequently we cannot know the precise criterion that each allocator optimizes--but we can be certain that it is not the makespan, since virtual pages located far from each other may be mapped to contiguous physical ones and vice versa (i.e. contiguous virtual pages can be mapped to non-neighboring physical ones). There is thus no point in trying to restrict the range of virtual addresses used.

There is quite a point, however, in trying to restrict the overall memory usage--or to minimize physical memory fragmentation. So the question is, in the context of the representation we are constructing, what could fragmentation look like? Our proposed answer is indicated by the two shaded rectangles in Figure \ref{fig:frag}. Recall that one description of fragmentation is "memory wastage"; the shaded areas are like gaps in a Tetris game. They represent segments which the allocator left unused, thus reserving higher addresses in order to handle all requests.

One might judge our formulation as too strict, since a non-moving allocator could not break job B in two and slide the left part down to cover the top fragmented area. Two points must be raised here: fragmentation is partly defined by the program's behavior, and it thus makes sense for portions of it to be inevitable. Moreover, what matters most is 2DBP itself. Computations performed on it, fragmentation included, are secondary. This statement does not mean to devalue fragmentation as a phenomenon--such a stance would go against our own motivation. It just stresses the importance of \textit{first} establishing a useful substrate. In our paper, fragmentation plays the crucial role of bootstrapping 2DBP in the sense of a 2DBP-derived signal correlating with the real world. But again, nothing else must be considered more primary than the representation itself.

\section{Proposed Method}
\label{sec:meth}
\begin{figure}[H]%
    \centering
    \includegraphics[width=\columnwidth]{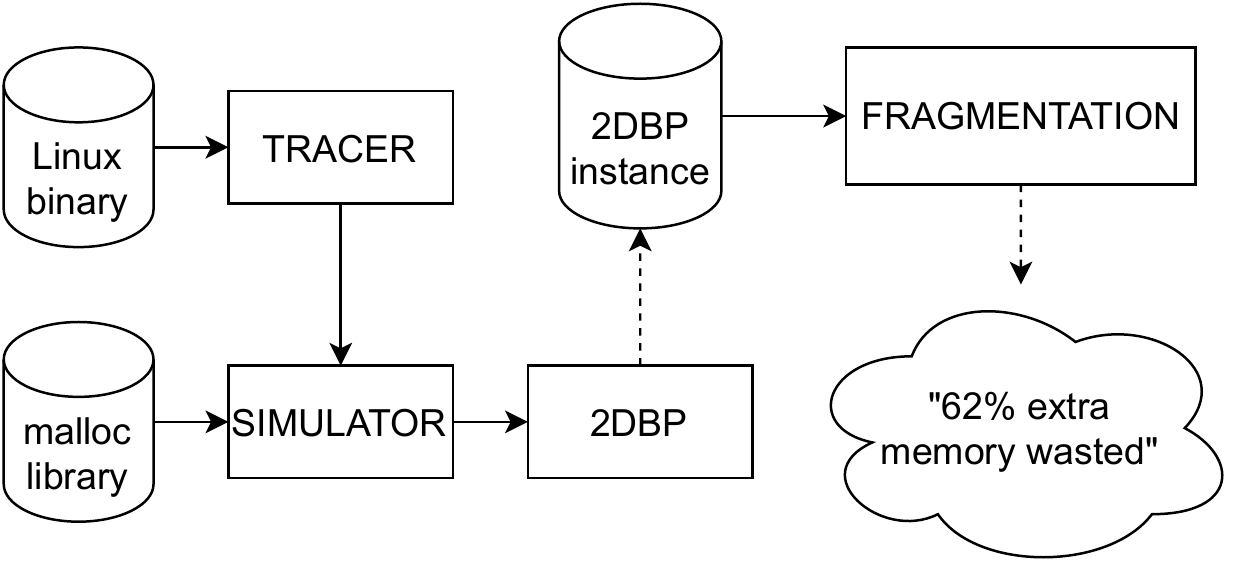}
    \caption{An overview of our method to produce 2DBP representations and to compute their fragmentation. Bold arrows are inputs and dotted arrows are outputs.}%
    \label{fig:over}%
\end{figure}

Our goal is to represent arbitrary pairs of Linux binaries and \texttt{malloc} implementations as 2DBP instances. An overview of our method is shown at Figure \ref{fig:over}. Inspired by~\cite{wilson1995surv} and ~\cite{fragsolved} we aimed for trace-based simulation. We focused on single-threaded programs to ensure reproducible behavior, and to simplify the simulation procedure.

\subsection{2DBP construction}
We log all of a program's calls to allocation functions. The resulting trace, along with the \texttt{malloc} implementation of interest, feeds our simulation module. The 2DBP component produces the final representation. Our architecture is modular to enable optimizations in each stage, since it must eventually handle realistic workload sizes.

\subsubsection{Requests tracing}
\label{sec:trace}
A reasonable question is why did we not leverage existing solutions such as \texttt{mtrace}\footnote{\href{https://linux.die.net/man/3/mtrace}{https://linux.die.net/man/3/mtrace}}, \texttt{heaptrack}\footnote{\href{https://github.com/KDE/heaptrack}{https://github.com/KDE/heaptrack}}, or tracing capabilities built in 
\texttt{malloc} implementations. Our decision was driven by the below points:

\begin{itemize}
    \item \texttt{mtrace} demands that the program of interest be modified so as to initialize the tool, while access to the application source code may not be feasible in practice.
    \item \texttt{heaptrack} and similar alternatives are, to begin with, extra dependencies on their own, which the user may want to avoid.
    \item existing alternatives impose larger overheads, since they store additional data (stack traces, call site addresses, etc.)
\end{itemize}

Our tracer is required to be \textit{complete}, catching allocations and deallocations all across the program's call stack. It must also be \textit{non-intrusive}, that is to imply zero actions regarding code instrumentation and compilation. It finally needs to be \textit{correct}: logged calls should belong to the traced program only, and not be polluted by dynamic memory operations of the tracer itself. To satisfy these requirements we target typical Linux processes forking no children. We also make use of several Linux and GNU utilities reported in the following paragraphs. Our mechanism is general enough to operate on any program in this context, from command line tools to application virtual machines.

\begin{figure}[t!]%
    \centering
    \includegraphics[width=\columnwidth]{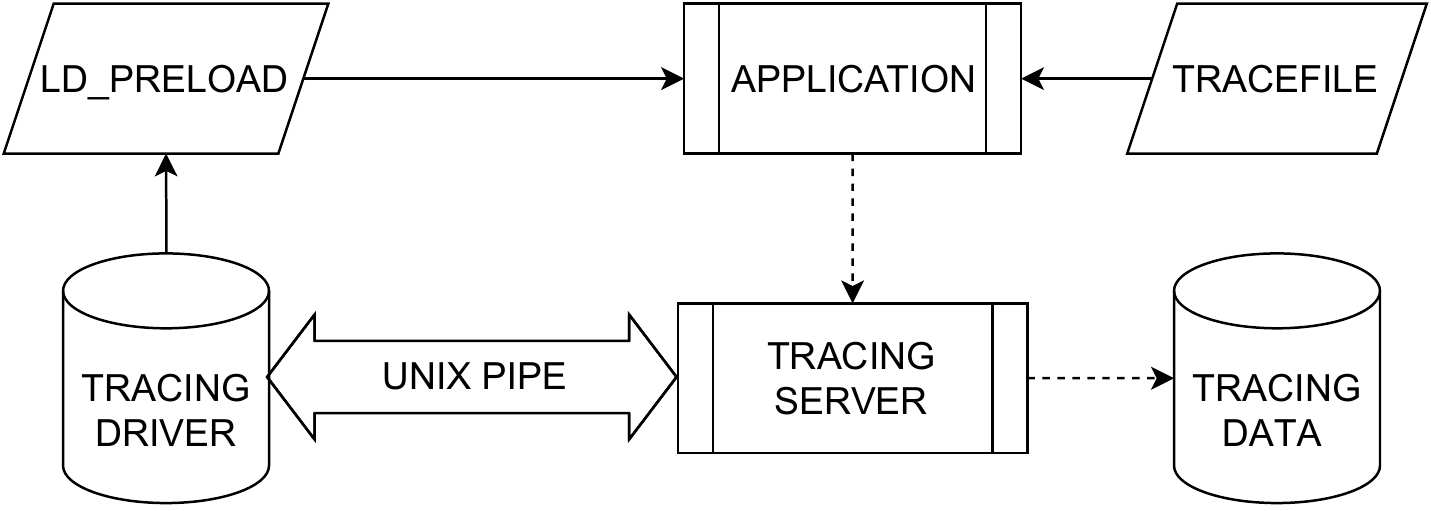}
    \caption{The tracing mechanism.}%
    \label{fig:trc}%
\end{figure}

The solution is sketched in Figure \ref{fig:trc}. The tracer is a shared library ("tracing driver") employing \texttt{dlsym}\footnote{\href{https://man7.org/linux/man-pages/man3/dlsym.3.html}{https://man7.org/linux/man-pages/man3/dlsym.3.html}} to interpose calls to \texttt{malloc}, \texttt{free}, \texttt{calloc}, \texttt{realloc}, \texttt{posix\_memalign}, \texttt{memalign}, \texttt{aligned\_alloc}, \texttt{pvalloc} and \texttt{valloc} during runtime. These were selected according to GNU's guidelines on replacing \texttt{malloc}\footnote{\href{https://www.gnu.org/software/libc/manual/html_node/Replacing-malloc.html}{https://www.gnu.org/software/libc/manual/html\_node/Replacing-malloc.html}.}. Beyond interposing the allocation interface, our tracer spawns a new process ("tracing server") which writes the actual logs to a CSV file. Tracing proceeds asynchronously via a Unix pipe of fixed capacity. The application makes a request, invoking our library driver's interposing code. The request is materialized via the true function obtained from \texttt{dlsym}. Before responding to the running program, the driver informs the server through the pipe. It then returns, allowing normal application execution to proceed. As long as there remain unread messages in the pipe, our server consolidates a CSV file containing the tracing data. The structure of the stored tracing data is shown at Table \ref{tab:trc}.

\begin{table}[t!]
    \centering
    \caption{Trace file structure. The \texttt{els\_num} field is used for tracing \texttt{calloc}, which returns a number of elements, each element of a certain size.}
    \label{tab:trc}
    \begin{tabular}{|c|c|c|c|}
        \hline
        \textbf{CSV Field} & \textbf{Request 1} & \textbf{Request 2} & \textbf{Request 3} \\ \hline
        \textbf{req\_type} & \texttt{malloc} & \texttt{free} & \texttt{calloc} \\ \hline
        \textbf{in\_address} & \texttt{(nil)} & \texttt{0x55A} & \texttt{(nil)} \\ \hline
        \textbf{out\_address} & \texttt{0x55A} & \texttt{(nil)} & \texttt{0x63B} \\ \hline
        \textbf{el\_size} & \texttt{12} & \texttt{(nil)} & \texttt{128} \\ \hline
        \textbf{els\_num} & \texttt{1} & \texttt{(nil)} & \texttt{1000} \\ \hline
    \end{tabular}

\end{table}

\subsubsection{Placement simulation}
\label{sec:sim}
2DBP perceives only two kinds of requests, namely allocation of $n$ bytes and deallocation of occupied memory. But a real trace file may include operations with more complex semantics, such as \texttt{calloc}. We thus unpack all calls to combinations of the two elementary operations, \texttt{malloc} and \texttt{free}. The counterargument to address is the proposed unpacking's effect on original program behavior. A short yet concise answer is that if along our course we distorted program behavior more than we should, no connection with RSS would have been uncovered. The unpacking scheme is described in Table \ref{tab:unpck}.

\begin{table}[H]
    \centering
    \caption{Rules for unpacking request traces to \texttt{malloc} and \texttt{free} operations. \texttt{s} stands for "size", \texttt{p} for "pointer", \texttt{n} for "number", \texttt{a} for "alignment".}
    \label{tab:unpck}
    \begin{tabular}{|c|c|}
        \hline
        \textbf{Original Operation} & \textbf{Transform} \\ \hline
        \texttt{malloc(s)} & \texttt{malloc(s)} \\ \hline
        \texttt{free(p)} & \texttt{free(p)} \\ \hline
        \texttt{calloc(s,n)} & \texttt{malloc(n$*$s)} \\ \hline
        \texttt{realloc(p,s)} & \texttt{free(p); malloc(s)} \\ \hline
        \texttt{posix\_memalign(p,a,s)} & \texttt{malloc(s)} \\ \hline
        \texttt{aligned\_alloc(a,s)} & \texttt{malloc(s)} \\ \hline
        \texttt{valloc(s)} & \texttt{malloc(s)} \\ \hline
        \texttt{memalign(a,s)} & \texttt{malloc(s)} \\ \hline
        \texttt{pvalloc(s)} & \texttt{malloc(s)} \\ \hline
    \end{tabular}
\end{table}

\begin{figure*}[t!]
    \centering
    \includegraphics[width=\textwidth]{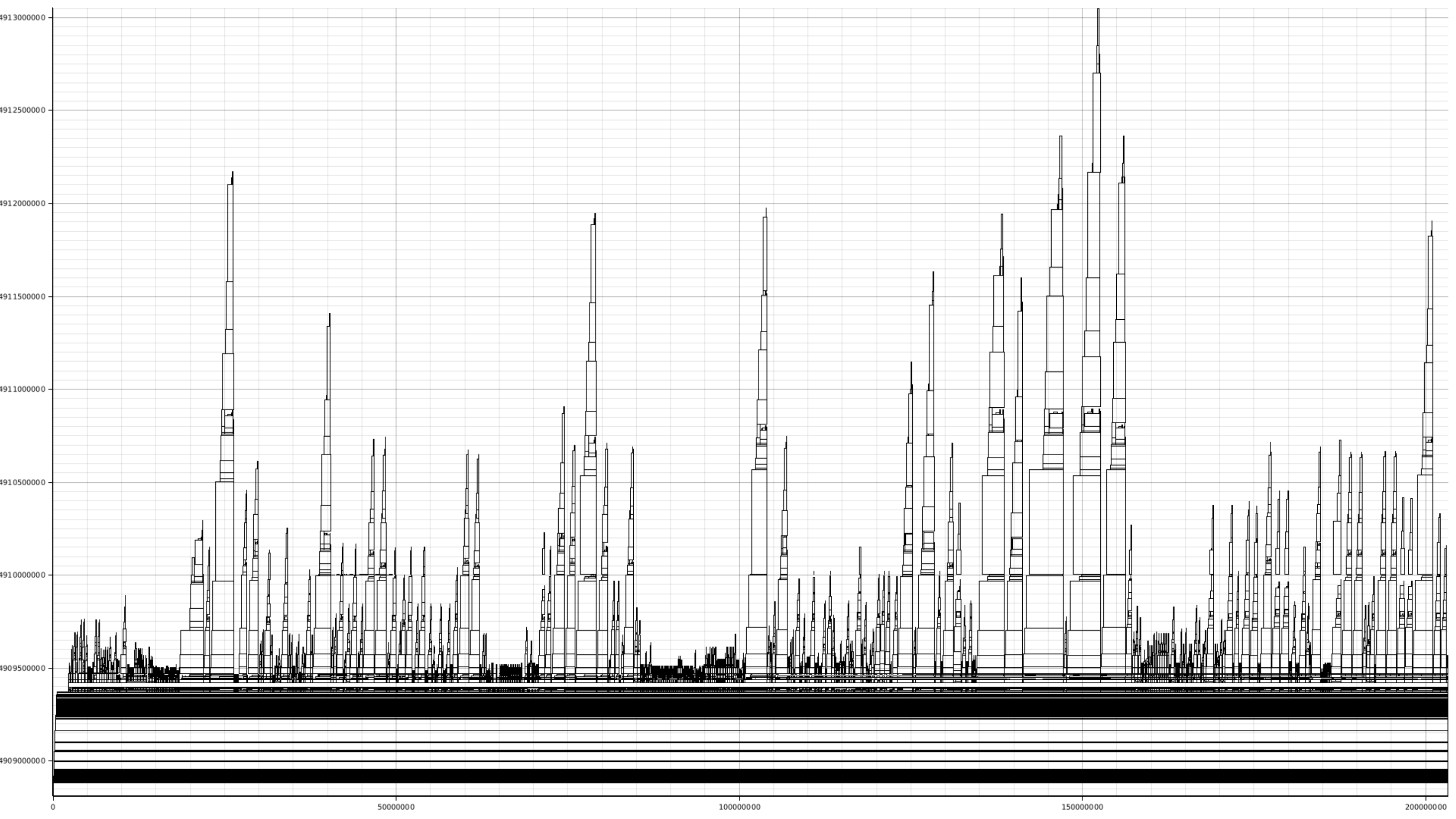}
    \caption{2DBP representation of the \texttt{node-express-loadtest} benchmark linked to \texttt{glibc}. The horizontal axis is time in bytes allocated. The vertical axis is virtual addresses. Around 270,000 rectangles are depicted, hence the solid black areas. \texttt{glibc} spawned 9 more mappings while serving this workload.}
    \label{fig:nodeserver}
\end{figure*}

For an accurate simulation of an allocator's placement decisions to a series of requests, several additional factors have to be taken into account:

\begin{itemize}
    \item placements must not be polluted by dynamic memory operations of the simulator itself.
    \item each job's height must reflect the actual amount of memory returned by the allocator. Block layout and segregated fit policies lead to jobs of \textit{different size than the one requested.} This difference amounts to internal fragmentation, which we have already put outside this paper's scope.
    \item memory mappings created by the allocator via \texttt{mmap}\footnote{\href{https://man7.org/linux/man-pages/man2/mmap.2.html}{https://man7.org/linux/man-pages/man2/mmap.2.html}} must be recorded, and used to annotate each request's product. Few--if any--modern allocators make exclusive use of the \texttt{sbrk}-backed\footnote{\href{https://linux.die.net/man/2/sbrk}{https://linux.die.net/man/2/sbrk}} heap space.
    \item memory leaks have to be treated.
    \item time progression, measured in allocated bytes, has to be maintained.
\end{itemize}

Policy simulation does not reproduce the original program's RSS waveform, since no memory access information is stored during the tracing stage. 2DBP lives entirely in virtual, not physical, memory. This works to our advantage, since it enables us to examine the extent to which events in virtual memory affect real-world performance.

\begin{figure}[t!]%
    \centering
    \includegraphics[width=\columnwidth]{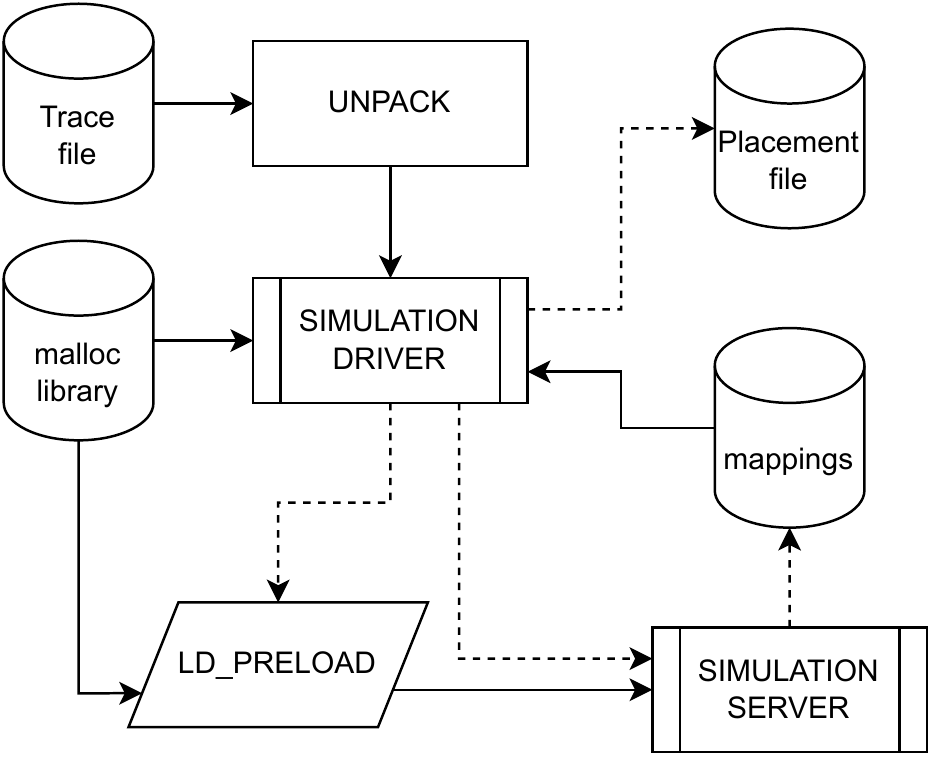}
    \caption{Policy simulation forms our method's most complex module.}%
    \label{fig:sim}%
\end{figure}

Our simulation procedure is depicted on Figure \ref{fig:sim}. Its backbone is the process annotated as "simulation driver". To ensure placement correctness, the driver spawns a clean new process ("simulation server") via \texttt{fork}\footnote{\href{https://man7.org/linux/man-pages/man2/fork.2.html}{https://man7.org/linux/man-pages/man2/fork.2.html}} and \texttt{execve}\footnote{\href{https://man7.org/linux/man-pages/man2/execve.2.html}{https://man7.org/linux/man-pages/man2/execve.2.html}}. Driver/server communication uses two Unix pipes of fixed capacity, similar to the tracing stage. The driver scans the next request, sends it to the server, and waits for its response. The server makes the actual request and responds back. The cycle is repeated in lockstep until all requests have been scanned.

To record correct block sizes, the server uses the values returned by \texttt{malloc\_usable\_size}\footnote{\href{https://man7.org/linux/man-pages/man3/malloc_usable_size.3.html}{https://man7.org/linux/man-pages/man3/malloc\_usable\_size.3.html}}. If the simulated allocator includes metadata in its block layout (like the GNU implementation does), this is also taken into account. Memory mappings are consulted by the simulation driver via the process-specific \texttt{/proc/[PID]/maps}\footnote{\href{https://man7.org/linux/man-pages/man5/proc.5.html}{https://man7.org/linux/man-pages/man5/proc.5.html}} file.

The driver keeps a list of live memory allocated by the server. When the last request has been scanned, jobs remaining in the list correspond to leaked memory. We update the respective entries in the produced placement file accordingly. Time is initialized and kept by the driver. It is updated if the last request scanned was a \texttt{malloc}. The final placement data is structured as a CSV file carrying the below fields:

\begin{itemize}
    \item \textbf{job\_id}: a unique identifier of a memory block that remained allocated for a specific amount of time, also known as a job.
    \item \textbf{block\_size}: the height of the job.
    \item \textbf{t\_start}: the point in time when the \texttt{malloc} call creating the job was done.
    \item \textbf{t\_end}: the point in time when the corresponding block was freed. All jobs representing leaked memory share the same value in this field, namely the total number of bytes allocated.
    \item \textbf{address}: the virtual address of the block's first byte, or where the block was placed.
    \item \textbf{map\_start}: the start address of the memory mapping to which the job belongs.
\end{itemize}

\subsubsection{Final representation}
A 2DBP instance is a \textit{set of ordered subsets}: each subset contains jobs belonging to the same memory mapping. Subsets are ordered in ascending job creation time. Ordering the subsets (i.e. the mappings) themselves makes no sense. 
Given a placement file we split it into as many parts as there are mappings. Each mapping can be treated independently--though all jobs in all mappings are indirectly bound together in the temporal dimension. To assist intuition and data visualization, we normalize each mapping's jobs in two ways: we set the first job's creation time to zero, and also set the lowest-placed job's address to zero. Individually normalized mappings and their jobs constitute our final 2DBP representation. Figure \ref{fig:nodeserver} displays an example taken from our experimental data.

\subsection{Fragmentation}
\label{sec:frag}
As discussed in Section \ref{sec:bck}, we view fragmentation's source to be the gaps between 2DBP jobs. To measure it we choose the most intuitive formulation: \textit{a workload-allocator pair's fragmentation is the sum of its gaps divided by the sum of its jobs' areas.} We compute fragmentation across all the $M$ mappings spawned by a workload-allocator pair via Equation \ref{eq:frag}:

\begin{equation}
    F_T = \frac{\sum_{i=1}^{M} F_{mi}}{\sum_{i=1}^{M} L_{mi}}
    \label{eq:frag}
\end{equation}

\begin{figure}[t!]%
    \centering
    \includegraphics[width=\columnwidth]{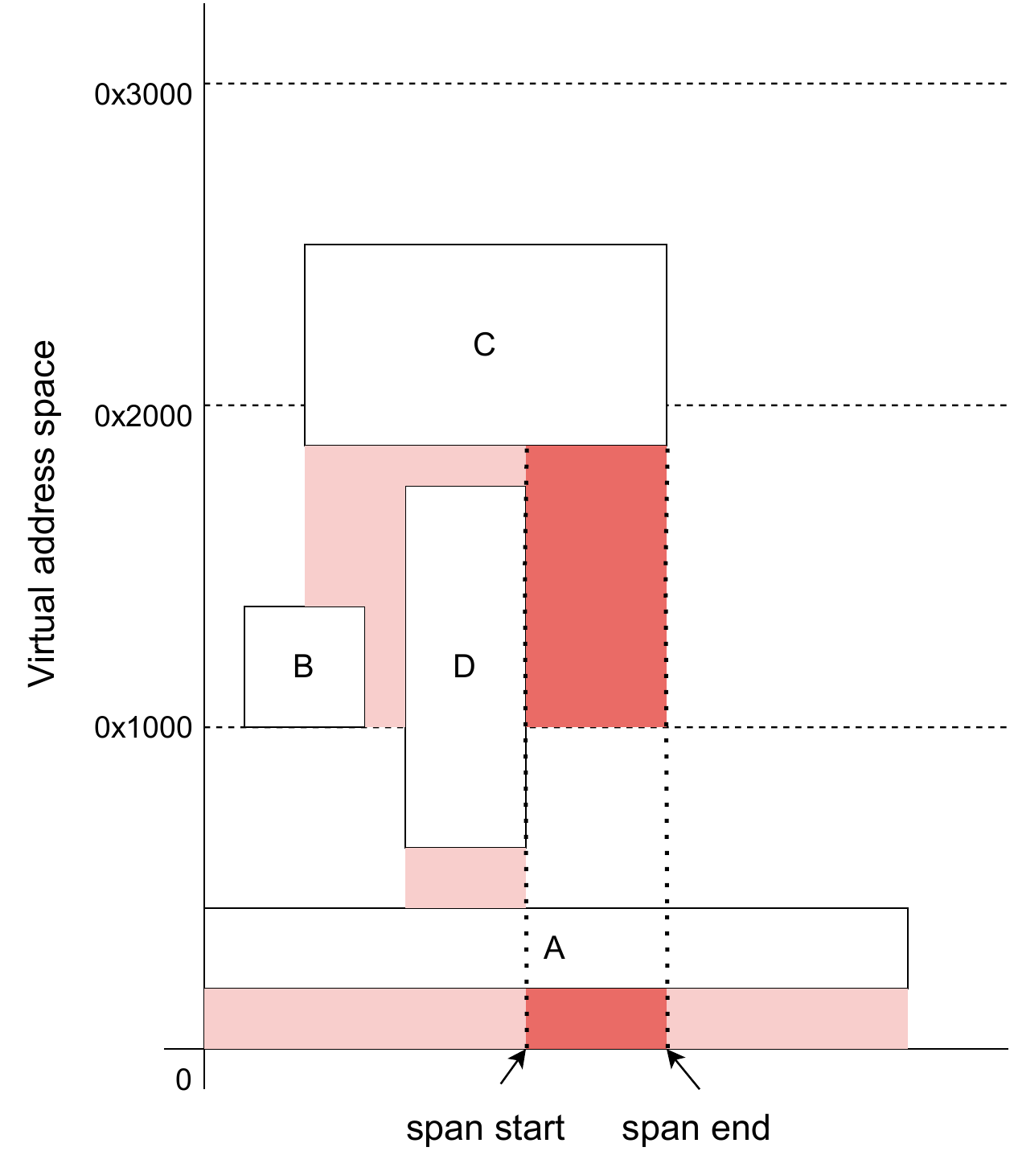}
    \caption{Gap identification algorithm. Axes are identical to those of Figure \ref{fig:frag}. Horizontal dashed lines are page boundaries. The current span holds 2 areas: one starting at address 0x000 and containing only job A (D belongs to the same page but is not live during the current span) and another starting at address 0x1000 and containing only job C.}%
    \label{fig:falgo}%
\end{figure}

The term $F_{mi}$ is derived by summing the areas of gaps between each mapping's jobs. $L_{mi}$ sums the areas of the jobs themselves. We illustrate our algorithm in Figure \ref{fig:falgo}: gaps between jobs are shown as lightly and darkly shaded areas.

Our plot is drawn \textit{in medias res}--lightly shaded areas were and will be computed in previous and future iterations, while the two darkly shaded ones are captured by the present iteration. To this we focus. It involves a vertical slice that we call a \textit{span}. Spans are delimited by job beginnings and endings. Within them nothing new happens; thus they can be traversed \textit{vertically} for new gaps to be found. 

Virtual page boundaries are drawn as horizontal dashed lines. We do not allow gaps to cross those boundaries, since there is no guarantee of maintained contiguity between virtual and physical memory. Gaps must always have a same-page job as their ceiling. This puts more pressure on the allocator's placement decisions and discounts the effect of limitations it cannot overcome.

\begin{figure}[t!]%
    \centering
    \includegraphics[width=\columnwidth]{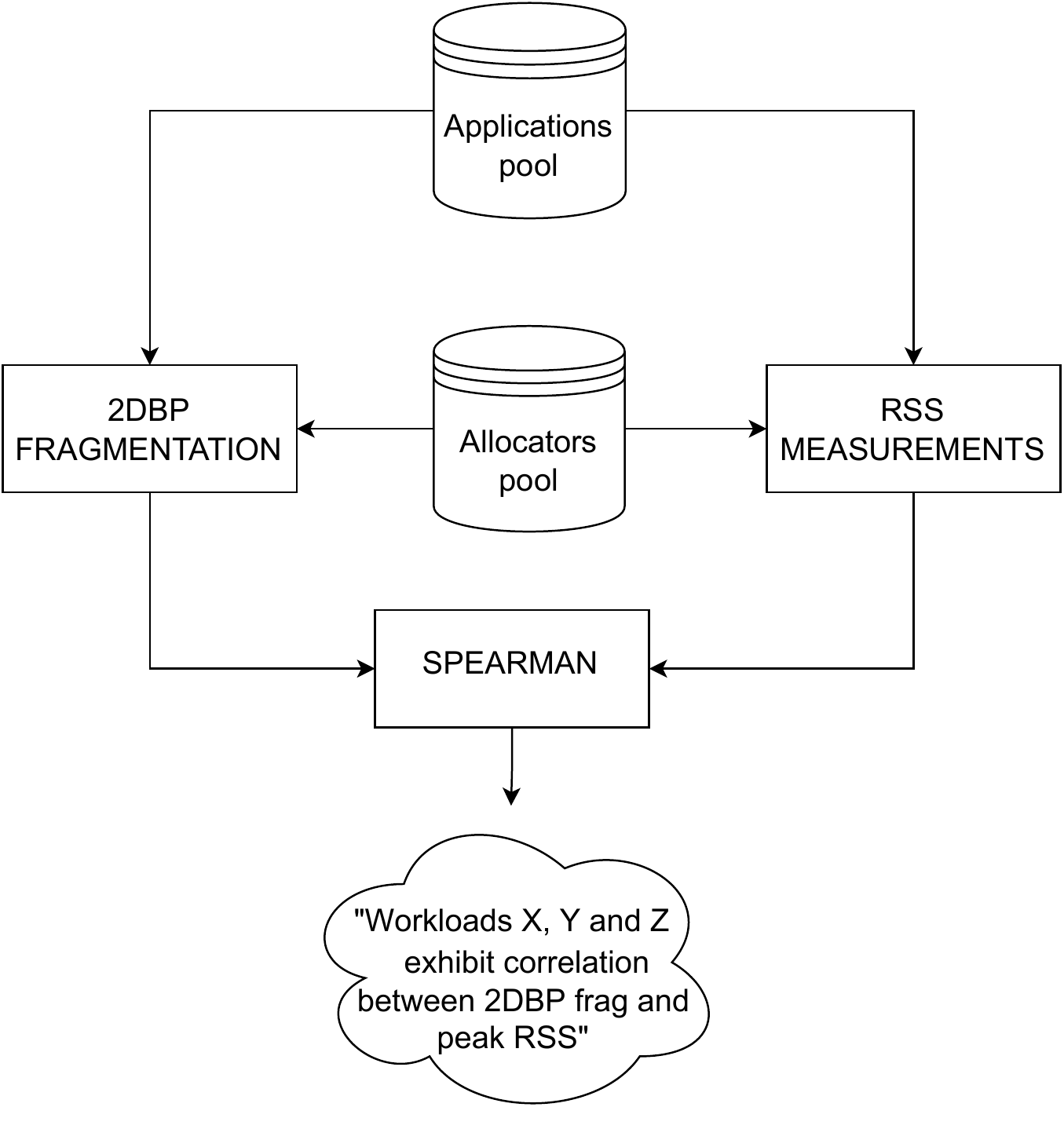}
    \caption{Our experimental procedure for investigating 2DBP's connection to peak RSS.}%
    \label{fig:charalgo}%
\end{figure}

\section{Evaluation}
\label{sec:res}
Up to this point, we have proposed a methodology that allegedly captures workload-allocator interaction. To evaluate our claim, a connection between our representation and a valuable physical memory-based measure must be made. We select maximum RSS as our target measure, and form a hypothesis around it: the cost of high fragmentation is most evident at the moment of highest memory usage~\cite{wilson1995surv}, i.e. at peak RSS. If 2DBP actually captures workload-allocator interaction, then computing fragmentation on it yields a good approximation of real\footnote{Recall that the hardness and ambiguity of measuring real fragmentation was this paper's starting point.} fragmentation. Consequently, 2DBP-based fragmentation correlates with peak RSS \textit{if and only if 2DBP as a whole is a valid representation.}

\subsection{Experimental setup}

The correlation we are looking for is a monotonically increasing function; we expect higher fragmentation to cause higher peak RSS. Thus we test our hypothesis via the process depicted in Figure \ref{fig:charalgo}. We collect data from a pool of applications linked to a pool of state-of-the-art allocators. We then compute workload-specific Spearman correlation coefficients for the peak RSS and fragmentation measurements.

Table \ref{tab:setup} lists the machines and allocators used in our experiments. Table \ref{tab:summary} provides a concise summary of benchmarks used as well as measurements recorded. All benchmarks come from the Single-Threaded Tests collection on openbenchmarking.org\footnote{\href{https://openbenchmarking.org/suite/pts/single-threaded}{https://openbenchmarking.org/suite/pts/single-threaded}}. We used the Phoronix Test Suite\footnote{\href{https://www.phoronix-test-suite.com/}{https://www.phoronix-test-suite.com/}} to install and run the applications. The Linux command \texttt{time}\footnote{\href{https://man7.org/linux/man-pages/man1/time.1.html}{https://man7.org/linux/man-pages/man1/time.1.html}} was used for memory footprint measurements. We took 10 samples per workload-allocator pair and calculated mean values and standard deviations.

\begin{table}[t!]
    \centering
    \caption{List of materials used for evaluation.}
    \label{tab:setup}
    \begin{tabular}{|c|c|c|}
        \hline
        \multicolumn{3}{|c|}{\textbf{Machines}} \\ \hline
        \textbf{Machine} & \textbf{Spec} & \textbf{Value} \\ \hline
        \multirow{7}{*}{A} & Cores & 12 \\ \cline{2-3}
        & Clock frequency & 2.2 GHz  \\ \cline{2-3}
        & Main memory & 16 GiB \\ \cline{2-3}
        & L1i cache & 192 KiB \\ \cline{2-3}
        & L1d cache & 192 KiB \\ \cline{2-3}
        & L2 cache & 1.5 MiB \\ \cline{2-3}
        & L3 cache & 9 MiB  \\ \hline
        \multirow{7}{*}{B} & Cores & 8 \\ \cline{2-3}
        & Clock frequency & 3.4 GHz  \\ \cline{2-3}
        & Main memory & 32 GiB \\ \cline{2-3}
        & L1i cache & 128 KiB \\ \cline{2-3}
        & L1d cache & 128 KiB \\ \cline{2-3}
        & L2 cache & 1 MiB \\ \cline{2-3}
        & L3 cache & 8 MiB  \\ \hline
        \multirow{3}{*}{Both} & OS & Ubuntu 20.4 LTS \\ \cline{2-3}
        & Architecture & x86\_64  \\ \cline{2-3}
        & Page size & 4096 B \\ \hline
        \hline
        \multicolumn{3}{|c|}{\textbf{Allocators (common in both machines)}} \\ \hline
        \multicolumn{2}{|c|}{\textbf{Name}} & \textbf{Release} \\
        \hline
        \multicolumn{2}{|c|}{glibc} & 2.31\tablefootnote{\href{https://guix.gnu.org/packages/glibc-2.31/}{https://guix.gnu.org/packages/glibc-2.31/}} \\ \hline
        \multicolumn{2}{|c|}{jemalloc~\cite{jemalloc}} & 5.3.0\tablefootnote{\href{https://github.com/jemalloc/jemalloc/releases/tag/5.3.0}{https://github.com/jemalloc/jemalloc/releases/tag/5.3.0}} \\ \hline
        \multicolumn{2}{|c|}{mimalloc~\cite{mimalloc}} & 2.0.9\tablefootnote{\href{https://github.com/microsoft/mimalloc/releases/tag/v2.0.9}{https://github.com/microsoft/mimalloc/releases/tag/v2.0.9}} \\ \hline
        \multicolumn{2}{|c|}{snmalloc~\cite{snmalloc}} & 0.6.1\tablefootnote{\href{https://github.com/microsoft/snmalloc/releases/tag/0.6.1}{https://github.com/microsoft/snmalloc/releases/tag/0.6.1}} \\ \hline
    \end{tabular}
\end{table}

\subsection{Performance remarks}
Before proceeding to our main results, we note some data related to the performance of our tools in Table \ref{tab:perf}. These are merely indicative and do not intend to provide a complete view of the measures involved, since we observed that performance is heavily affected by each stage's input. For instance, the tracer's overhead is analogous to the traced program's request density, i.e. how many requests per second are made. Another example is fragmentation computation, where the jobs' topology in the 2DBP plane controls the resulting throughput; in a less dense occasion than the one reported, our tool managed to process 37 thousand jobs per second.

\begin{table}[t!]
    \centering
    \caption{Sample performance data.}
    \label{tab:perf}
    \begin{tabular}{|c|c|}
        \hline
        \multicolumn{2}{|c|}{\textbf{Requests tracing}} \\ \hline
        \texttt{heaptrack}'s overhead & 150\% \\ \hline
        Proposed mechanism's overhead & \textbf{94\%} \\ \hline
        Requests stored per MiB & 34K \\ \hline
        \hline
        \multicolumn{2}{|c|}{\textbf{Policy simulation \& 2DBP construction}} \\ \hline
        Throughput (requests/second) & 60K \\ \hline
        Jobs stored per MiB & 23K \\ \hline
        \hline
        \multicolumn{2}{|c|}{\textbf{Fragmentation computation}} \\ \hline
        Throughput (jobs/second) & 5K \\ \hline
    \end{tabular}
\end{table}

\begin{table*}[t!]
    \centering
    \caption{Measurements summary. Each row represents one workload. 28 workloads were studied in total. For each workload we record its number of jobs, and the peak RSS/fragmentation ranges as smallest-biggest measured value couples. To test for monotonicity between fragmentation and peak RSS, we also compute Spearman's correlation coefficient per each case. The data correspond to experiments on machine A, but machine B gave very similar results.}
    \label{tab:summary}
    \begin{tabular}{|c|c|c|c|c|c|}
    \hline
    \textbf{Application} & \textbf{Input} & \textbf{Jobs} & \textbf{Peak RSS range (MiB)} & \textbf{Fragmentation range (\%)} & \textbf{Spearman} \\
    \hline
    \multirow{3}{*}{PHPBench tests} & 0 & 4021506 & 17.8-18.6 & 47-65 & -0.51 \\ \cline{2-6}
    & 1 & 61506 & 17.7-18.6 & 45-62.3  & -0.49 \\ \cline{2-6}
    & 2 & 40021506 & 17.8-18.5 & 47-65.5 & -0.41 \\ \hline
    \multirow{3}{1.4in}{Node.js Express HTTP load test} & 0 & 649395 & 54-63.2 & 29.3-33.6 & 0.33 \\ \cline{2-6}
    & 1 & 270774 & 51.2-59.1 & 18.8-23.5 & 0.06 \\ \cline{2-6}
    & 2 & 2039657 & 63.4-72.9 & 34.2-40.6 & 0.19 \\ \hline
    \multirow{3}{1.4in}{\textbf{System libxml2 parsing benchmark}} & \textbf{0} & 1163821 & 5.5-6.1 & 37-56.3 & \textbf{0.89} \\ \cline{2-6}
    & \textbf{1} & 270774 & 5.6-6.2 & 32.4-56.4 & \textbf{0.92} \\ \cline{2-6}
    & \textbf{2} & 202521 & 5.4-6.1 & 37.1-60.2 & \textbf{0.89} \\ \hline
    \multirow{3}{1.4in}{dcraw digital photo encoder} & 0 & 169 & 82.3-85 & 0-0.02 & 0.39 \\ \cline{2-6}
    & 1 & 29 & 82.3-84.8 & 0-0.03 & 0.39 \\ \cline{2-6}
    & 2 & 63 & 82.2-84.8 & 0-0.03 & 0.39\\ \hline
    \multirow{3}{1.4in}{\textbf{tjbench libjpeg-turbo (de)compression tests}} & 0 & 1490 & 84.3-86.3 & 0.03-0.04 & 0.37 \\ \cline{2-6}
    & \textbf{1} & 10442 & 18.3-20.3 & 0.2-0.3 & \textbf{0.78} \\ \cline{2-6}
    & \textbf{2} & 34618 & 9.5-12 & 0.4-0.6 & \textbf{0.97} \\ \hline
    \multirow{3}{1.4in}{\textbf{eSpeak-NG speech synthesizer}} & \textbf{0} & 984 & 4.8-6.9 & 3.6-7.9 & \textbf{0.97} \\ \cline{2-6}
    & \textbf{1} & 984 & 4.1-5.9 & 3-6.8 & \textbf{0.72} \\ \cline{2-6}
    & \textbf{2} & 984 & 4-5.8 & 1.2-2.2 & \textbf{0.79} \\ \hline
    \multirow{3}{1.4in}{\textbf{Gzip compression benchmark}} & \textbf{0} & 610992 & 3-5.3 & 22.7-47.5 & \textbf{0.77} \\ \cline{2-6}
    & \textbf{1} & 22963 & 2.8-5.2 & 20.1-50.7 & \textbf{0.77} \\ \cline{2-6}
    & \textbf{2} & 161008 & 2.9-5.2 & 19.5-44.3 & \textbf{0.77} \\ \hline
    \multirow{3}{1.4in}{Optcarrot Ruby NES emulator} & 0 & 655462 & 94.8-100.3 & 25.4-30.6 & -0.93 \\ \cline{2-6}
    & 1 & 655463 & 94.8-98.7 & 26-30.1 & -0.94 \\ \cline{2-6}
    & 2 & 659763 & 94.9-98.8 & 19.5-29.6 & -0.91 \\ \hline
    \textbf{Java NIST SciMark 2.0} & N/A & 11338 & 37.3-40.8 & 16.7-19.2 & \textbf{0.66}\\ \hline
    Bullet physics engine & N/A & 79950 & 38.6-42.2 & 1.4-3.2 & -0.81 \\ \hline
    \textbf{NGINX web server benchmark} & N/A & 1480981 & 6.1-105.8 & 0.1-0.2 & \textbf{0.78} \\ \hline
    bork file encrypter & N/A & 12837 & 34.2-36.7 & 16.7-19 & 0.42 \\ \hline
    \hline
    \multicolumn{2}{|r|}{Workloads in total:} & 28 & \multicolumn{2}{|r|}{Spearman >0.65 in:} & \textbf{13 (46.4\%)} \\
    \hline
    \end{tabular}

\end{table*}

\begin{figure}[t!]%
    \centering
    \includegraphics[width=\columnwidth]{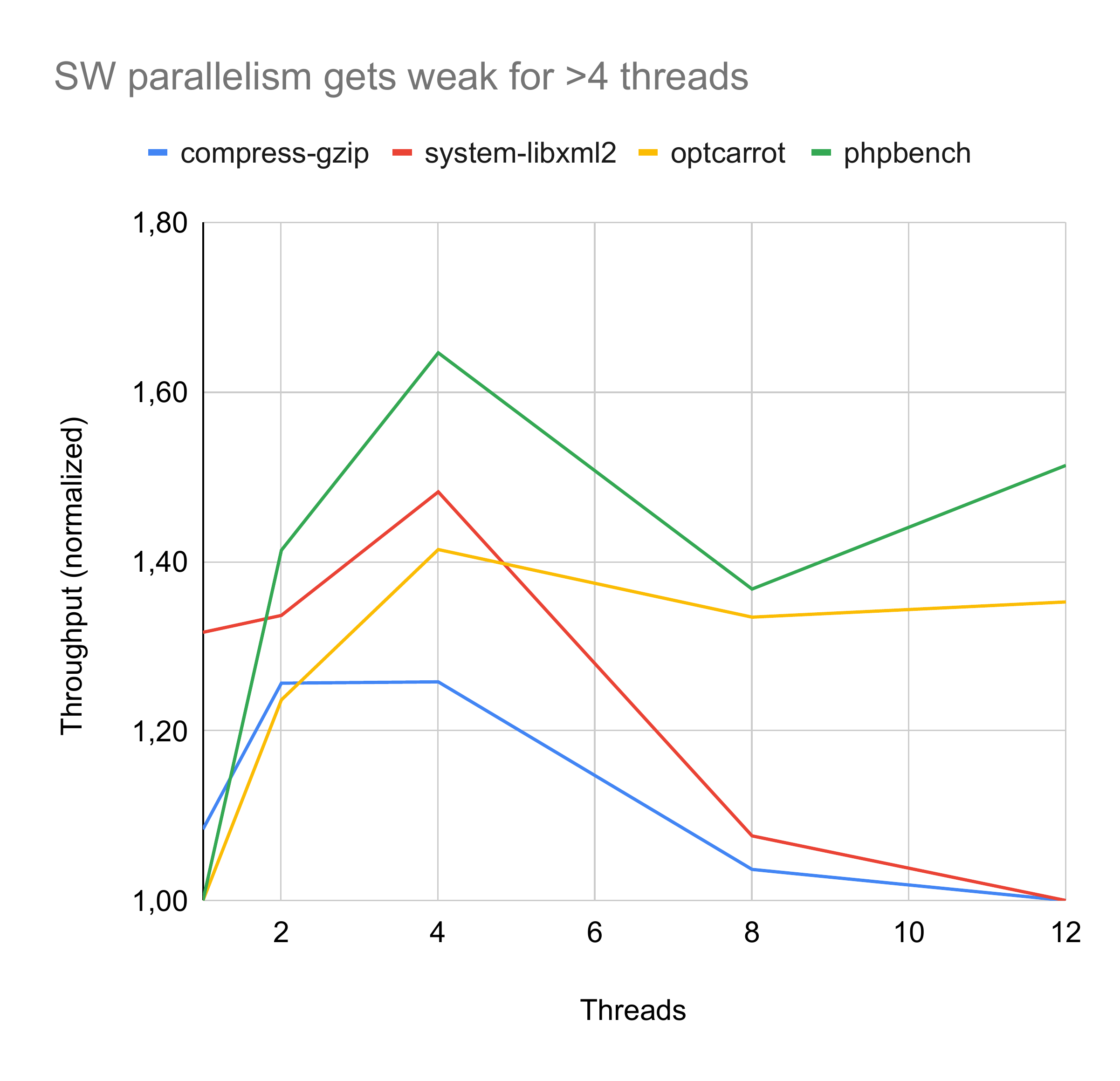}
    \caption{Parallelism gains in fragmentation computation performance (throughput as processed jobs per second) for 4 different workloads.}%
    \label{fig:threads}%
\end{figure}

We have made our tools as modular as possible to welcome future optimizations. The key takeaway from Table \ref{tab:perf} is the much smaller overhead of the tracer compared to \texttt{heaptrack}, a modern heap profiler. \texttt{heaptrack} records, as opposed to sampling approaches, all allocation calls, and thus allows for a fair comparison.

Figure \ref{fig:threads} depicts a brief investigation we conducted regarding fragmentation computation, an inherently parallel operation as discussed in Section \ref{sec:frag}. Most of the time, parallelism gains peak at 4 threads. We ascribe this lack of scalability to our present implementation, which has a very small computational kernel (traversing a span's areas and accumulating gaps found) and makes use of expensive atomic addition (the variable where all gaps are accumulated). These two factors combined make all threads finish their task very fast, and then stall until the last writing thread's atomic update of the accumulator variable is completed.

\begin{figure*}
     \centering
     \begin{subfigure}[t]{\textwidth}
         \centering
         \includegraphics[width=\textwidth]{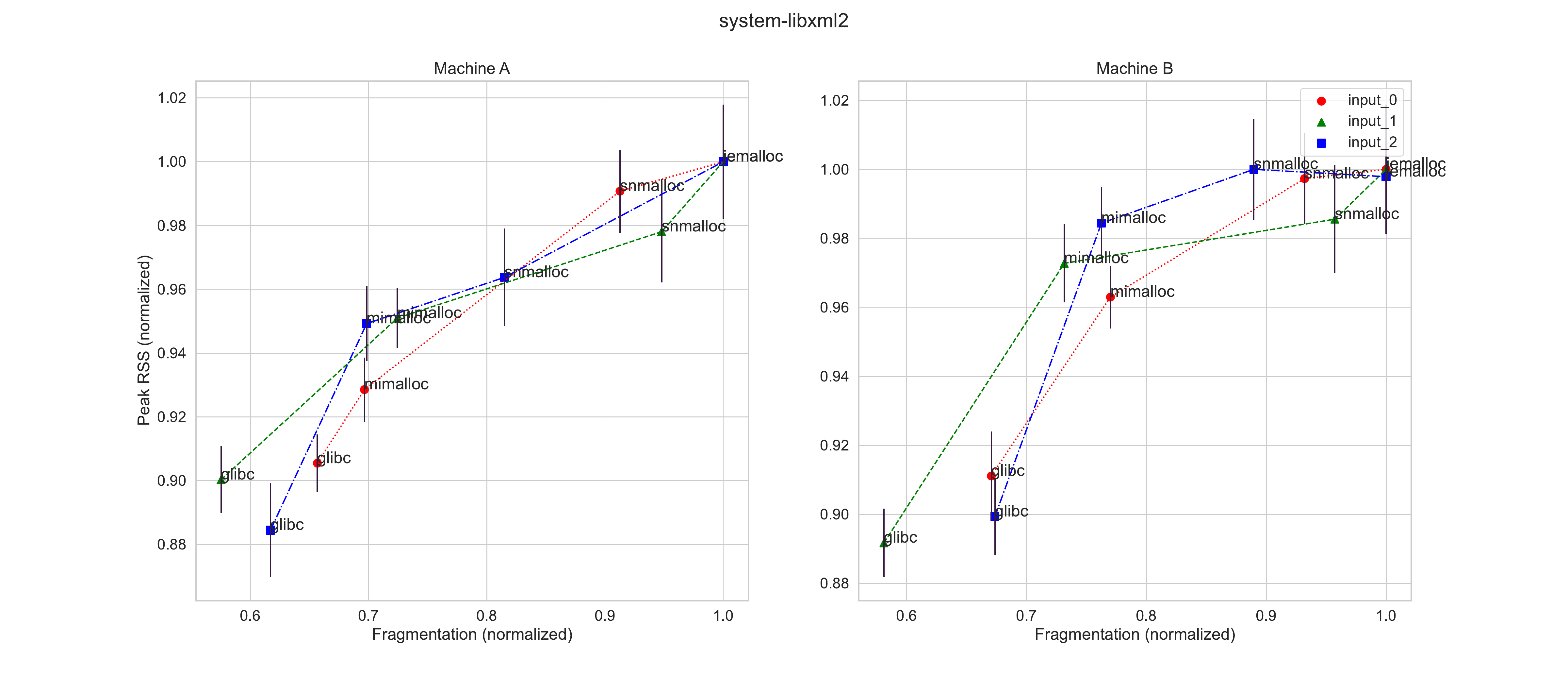}
         \caption{\texttt{system-libxml2}: lower fragmentation yields $\sim$10\% smaller memory footprint.}
         \label{fig:xml}
     \end{subfigure}
     \hfill
     \begin{subfigure}[b]{\textwidth}
         \centering
         \includegraphics[width=\textwidth]{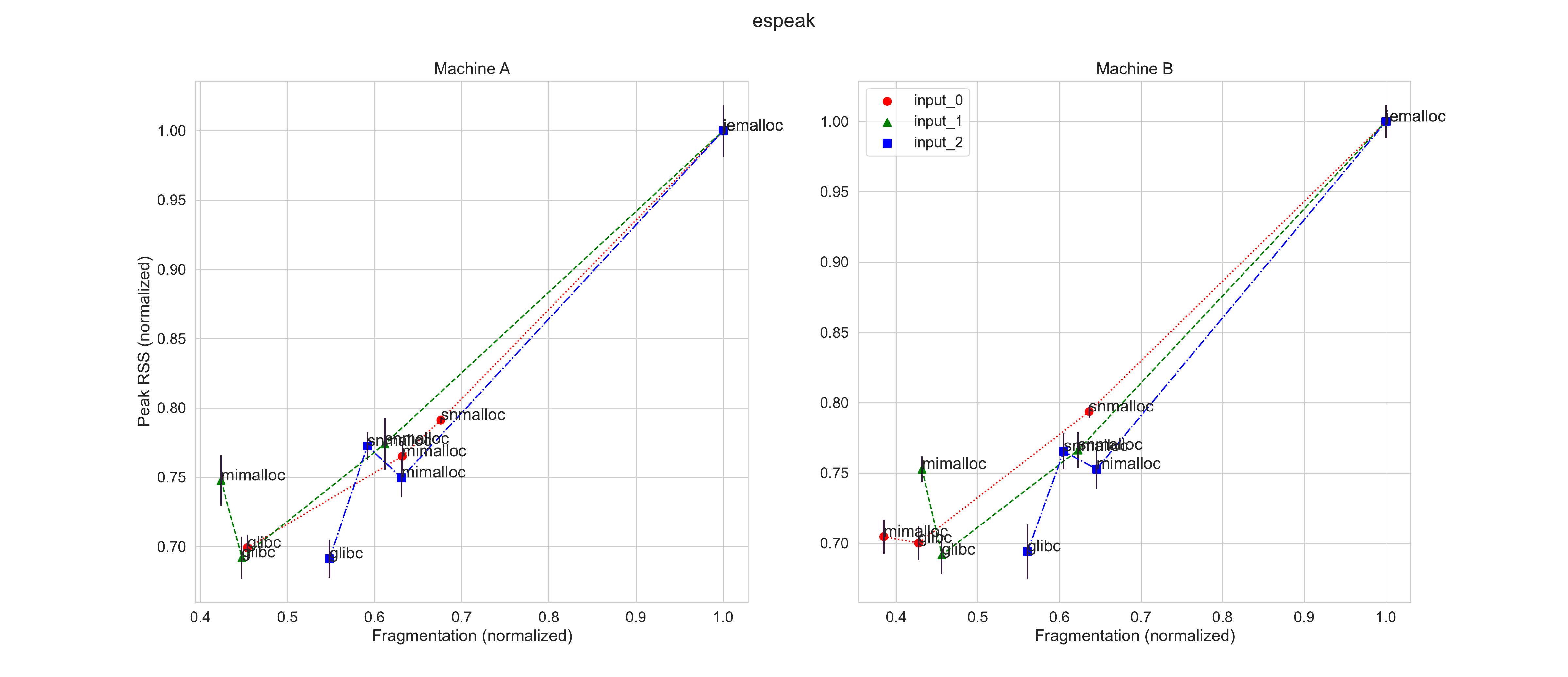}
         \caption{\texttt{espeak}: lower fragmentation yields $\sim$30\% smaller memory footprint.}
         \label{fig:espeak}
     \end{subfigure}
      \caption{Normalized scatter plots of peak RSS versus 2DBP-based fragmentation. With respect to peak RSS, each point is the mean value calculated after 10 executions--the black error bars correspond to the measurements' standard deviation.}
      \label{fig:mainres_1}
\end{figure*}

\begin{figure*}
     \centering
     \begin{subfigure}[t]{\textwidth}
         \centering
         \includegraphics[width=\textwidth]{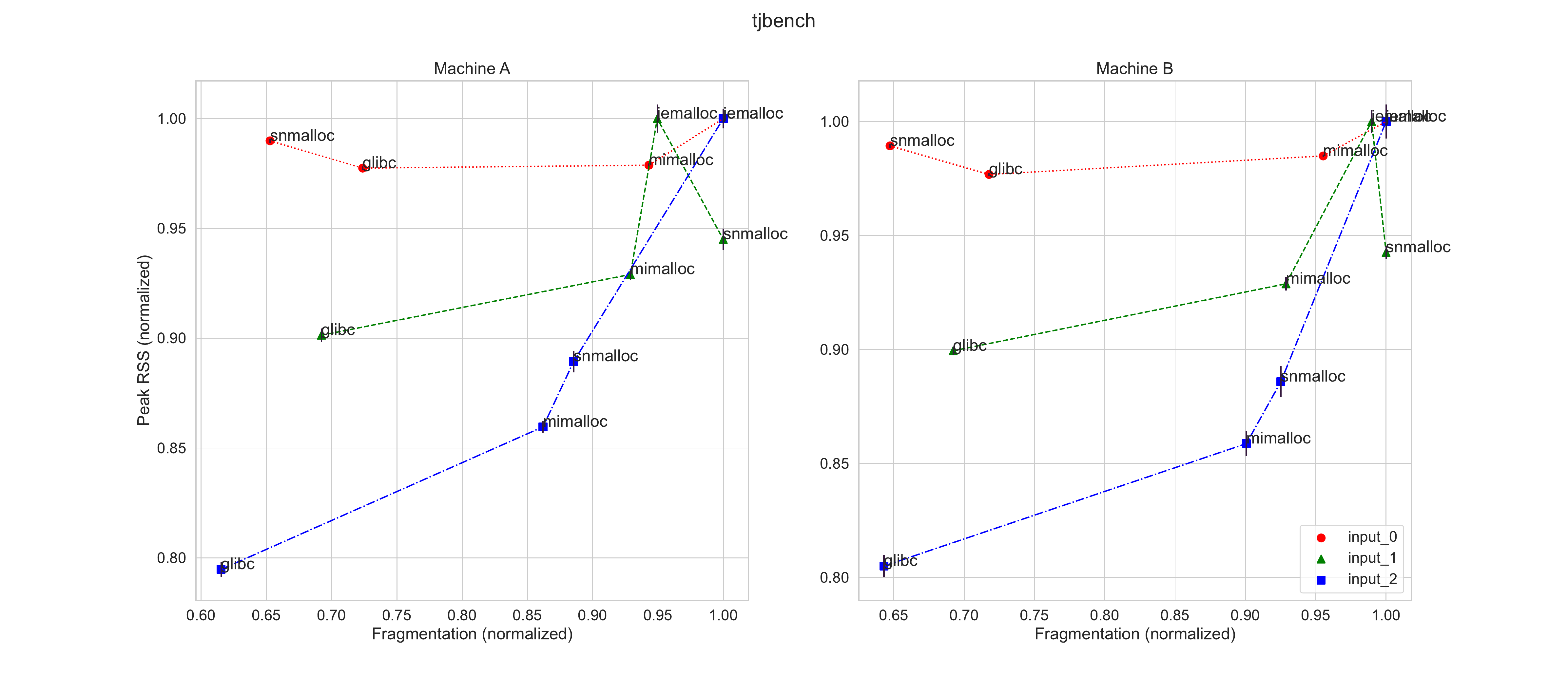}
         \caption{\texttt{tjbench}: lower fragmentation yields $\sim$20\% smaller memory footprint.}
         \label{fig:tjbench}
     \end{subfigure}
     \hfill
     \begin{subfigure}[b]{\textwidth}
         \centering
         \includegraphics[width=\textwidth]{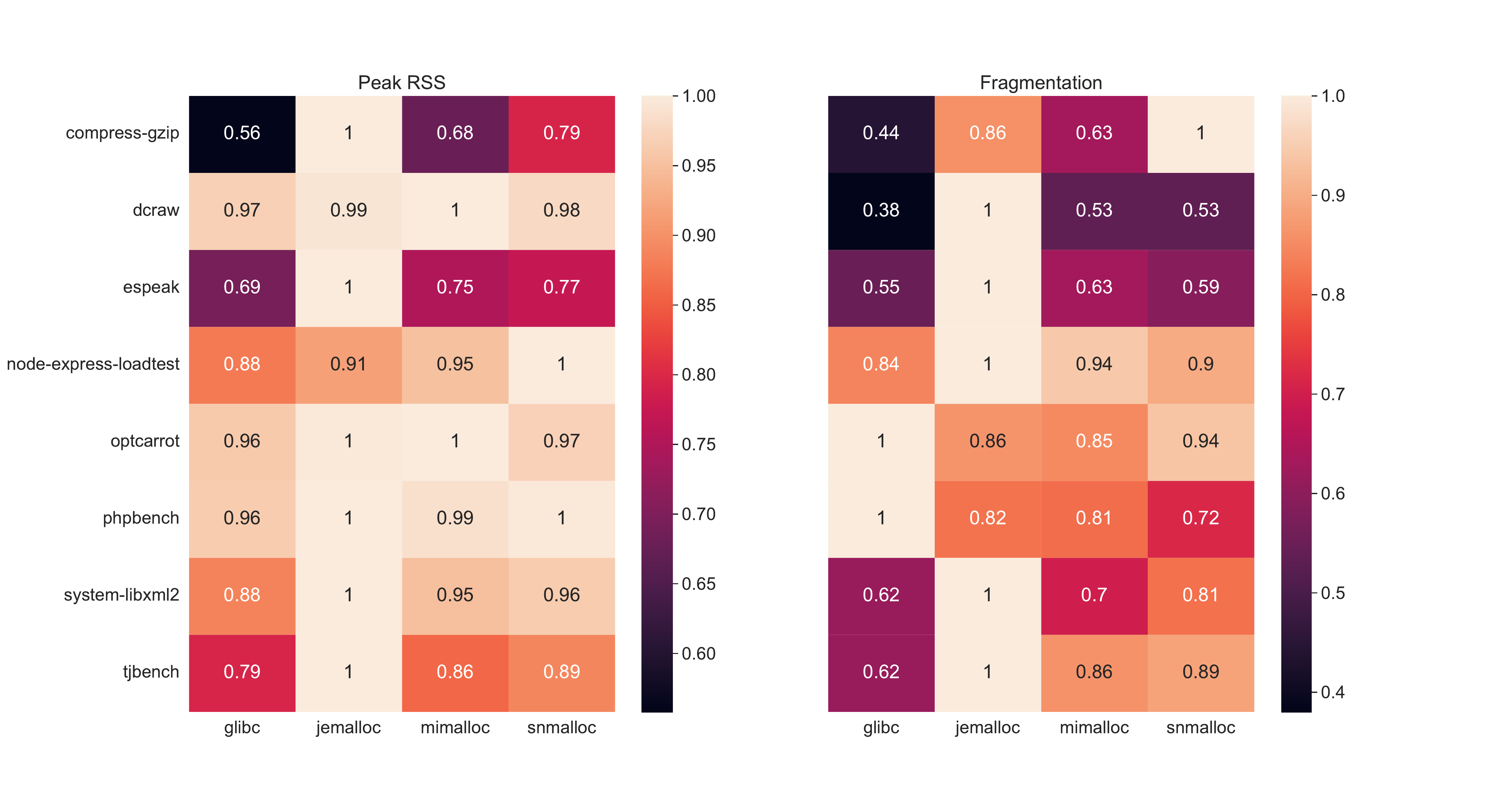}
         \caption{Normalized heatmaps for \texttt{input\_2}. Note how they can assist in characterizing allocators themselves: \texttt{jemalloc} often exhibits, for instance, the worst performance both in terms of peak RSS as well as fragmentation. The completely converse holds for \texttt{glibc}.}
         \label{fig:heat}
     \end{subfigure}
      \caption{One more scatter plot and a heatmap.}
      \label{fig:mainres_2}
\end{figure*}

\section{Discussion}
\label{sec:disc}
We now come to discuss our results and their potential implications.
\subsection{On the relationship between 2DBP-based fragmentation and maximum RSS}
\label{sec:mainarg}
The total amount of physical memory consumed by a workload is affected by many factors. Particularly with respect to dynamic memory allocation, 2DBP encodes the following:

\begin{itemize}
    \item what distribution do the sizes requested follow?
    \item in what patterns is memory freed by the application?
    \item what size classes does the allocator use?
    \item does the block layout include any metadata?
    \item what placement policy is followed?
    \item how do the previous points \textit{interact}?
\end{itemize}

All events our representation sees reside entirely in virtual memory, and even these events are a simplified version of the original workload's behavior. Thus our central research question: \textit{is such a restrained representation enough?} Our criterion must always be tied to physical memory. We want to examine whether actions taken in 2DBP are reflected on quantities of practical interest, like RSS. We want any conclusions derived from a 2DBP representation to be reliable.

Table \ref{tab:summary} gives us a first answer: in \textit{many} cases, our external fragmentation definition from Eq. \ref{eq:frag} does exhibit a monotonic relationship with peak RSS. We observe this in 13 out of 28 workloads ($46.4\%$) and 6 out of 12 programs ($50\%$), with the trends holding across two different machines. Gains from less fragmentation in 2DBP may reach up to $30\%$ smaller memory footprint, as is the case for Fig. \ref{fig:espeak}.

The expected correlation does not appear across all workloads due to factors affecting peak RSS which are not captured by 2DBP. Our representation disregards which specific function of the \texttt{malloc} family is used for allocation or reallocation. Memory access patterns to allocated memory remain invisible, and consequently so does the mapped/unmapped status of the virtual address space. Block indexing costs incurred by allocators are also not captured. Apparently, however, the interplay between \textit{captured} factors is often strong enough to overpower hidden information. 

\subsection{On potential implications}
\label{sec:impls}

2DBP is a product of oflline, trace-based simulation. The fact that it even partly validates our initial intuition, namely that peak RSS follows fragmentation, is hard to ignore. If the cause behind the presented evidence is mere chance, it is a rather unique, shapeshifting kind of chance. But if it is not, our correlation study must be viewed as a first empirical proof of 2DBP's informational potency.

In that case, we are obliged to contemplate not how our results ensure the robustness of this paper, but what their implications are for DSA as a field and a practice. Let us thus indulge in the following thought experiment: assume that we \textit{know} 2DBP to capture workload-allocator interaction. How does one capitalize on this knowledge? A first application would be identifying workloads that are provably sensitive to allocator policy--that is, workloads where significant savings in physical memory are expected if better placements are found. Figures \ref{fig:espeak} and \ref{fig:tjbench} provide some ready examples: there we see that reducing 2DBP-based fragmentation by $\sim 60\%$ and $\sim 40\%$ rewards us with $\sim 30\%$ and $\sim 20\%$ smaller peak RSS respectively. Such workloads would be perfect candidates for a benchmark suite evaluating placement policies.

Next, assume a sensitive workload that is to be executed on a memory-constrained machine. It is critical for the ones responsible to ensure that when it is deployed, the workload's peak RSS (or some other metric) is the minimum possible. A sandbox could be set up where different policies are iteratively tried on the workload's request trace, until the best one is found. The whole process would run offline, and not even access to the executable itself would be needed. Its request trace and a modifiable allocator would be the only required elements. 

The generation of (approximately) optimal placements with respect to some more relevant criterion than the classical makespan could also be studied. Lower bounds would then be assigned to sensitive workloads' achievable fragmentation. If the distance between said bounds and the top performing allocator were small, exploring custom policies for a particular workload would not be worth the effort. In the opposite case sandbox approaches like the one mentioned above could be explored.

Most importantly, 2DBP might yield more complex products: it could assist in performing feature extraction of workload-allocator pairs, for use in relevant machine learning tasks. We wonder what such tasks would look like; can, for instance, an allocator's policy be "learned"? Can similarity measures for allocators or workloads be established? The heatmap of Figure \ref{fig:heat} shows, for example, that \texttt{jemalloc} often has the worst performance in terms of both peak RSS and fragmentation. \texttt{glibc} exhibits the opposite behavior. We find great value in interpreting and refining such insights to assist better designs in the future.

\section{Related Work}
\label{sec:rw}
Wilson et al. have written the seminal treatment on DSA and the central role of fragmentation~\cite{wilson1995surv}. Johnstone and Wilson conduct the first study of RSS-based fragmentation definitions~\cite{fragsolved}. Berger et al. show that modern allocators perform acceptably well with respect to RSS-based fragmentation~\cite{reconsider}. Maas et al. propose a novel fragmentation definition incorporating chances of immediate memory reuse~\cite{realfrag}. Powers et al. and Maas et al. contribute notably unorthodox ways to deal with fragmentation~\cite{mesh, learnalloc}.

On the theoretical side Robson has computed worst case fragmentation bounds for the best fit and first fit placement policies~\cite{robson1977worst}. Optimal placement is reported as NP-hard by Garey and Johnson~\cite{garey_jognson_1979}. Chrobak and {\'S}lusarek formulate it as a 2DBP instance~\cite{chrobak1988some}. Buchsbaum et al. develop the state-of-the-art $\epsilon$-optimal algorithm for solving the general case with minimal makespan~\cite{buchsbaum}. Given our focus on 2DBP, we do not mention other formulations such as graph coloring~\cite{KIERSTEAD1991231}.

We did not find any works exploring representations of real workload-allocator interaction.

\section{Conclusion}
\label{sec:end}
This paper forms a connection between theoretical dynamic memory allocation and its real-world counterpart. It is motivated by a profound asymmetry between dynamic memory allocation's omnipresence and the scarcity of principled methods for understanding workload-allocator interaction. It describes a mechanism for extracting representations of workload-allocator pairs in the form of two-dimensional bin packing, and then proposes a novel fragmentation definition built on top. Despite operating on entirely virtual, simulation-generated data, our measure strongly correlates with a variety of workloads. Our study serves as a first piece of empirical evidence towards adopting bin packing-based methods for dynamic memory allocation.

\section{Enhancements and future research opportunities}
\label{sec:lims}
As discussed in Section \ref{sec:mainarg}, the information captured by 2DBP can be augmented. For instance, timestamps of first access can be added to the policy simulator's record, in order to divide job and gap bodies into "unmapped" and "mapped" portions--the unmapped parts would then be disregarded during fragmentation computation. Keeping the initial memory size requested by the program in a job's data and contrasting it with the size of the allocated block will make our method aware of internal fragmentation. Experiments focusing on a workload's working set size instead of peak RSS are an interesting path to follow. More complex, though less interpretable, metrics than Eq. \ref{eq:frag} can be derived via setting up appropriate regression tasks. Moreover, extending our method to parallel programs must be considered. We refrained from this only to ensure a stable, deterministic experimental setup, but there are works promising deterministic multithreaded behavior in the literature~\cite{dthreads}. 

\bibliographystyle{ACM-Reference-Format}
\bibliography{sample-base}

\appendix

\end{document}